\documentclass{jpsj-suppl}
\usepackage{txfonts} %Please comment out this line unless the txfonts package is availabe in your LaTeX system.
\usepackage{braket}
\usepackage{here}
\title{Neutrino Induced $^4$He Break-up Reaction \\ {\small -Application of the Maximum Entropy Method in Calculating Nuclear Strength Function-}}

\author{Tomoya \textsc{Murata}$^{1}$, Wataru \textsc{Horiuchi}$^{2}$, Toru \textsc{Sato}$^{1}$ and Satoshi X. \textsc{Nakamura}$^{1}$}

\inst{$^{1}$Department of Physics, Osaka University, Toyonaka, Osaka 560-0043, Japan \\
$^{2}$Department of Physics, Hokkaido University, Sapporo 060-0810, Japan}

\email{murata@kern.phys.sci.osaka-u.ac.jp}

\recdate{\today}

\abst{
 The maximum entropy method is examined as a new tool 
for solving the ill-posed inversion problem involved in the Lorentz integral transformation (LIT) method.
As an example, we apply the method to the spin-dipole strength function 
of $^4$He. We show that the method can be successfully used 
for inversion of LIT,  provided the
LIT function is available with a sufficient accuracy.}

\kword{neutrino reaction, supernova, Lorentz integral transformation method, 
maximum entropy method}

\begin{document}
\maketitle

\section{Introduction}

The neutrino-nucleus reactions play an important role for the heating 
and cooling mechanisms in the core collapse supernova explosion. 
Particularly the neutrino-$^4$He reactions have been of interest,
and their effects on the accelerating shock wave \cite{haxton1988} and 
nucleosynthesis \cite{fuller1995} have been studied.
The typical neutrino energy inside the supernova is tens of MeV, 
and the dominant reaction channel is $^4$He breakup.
Therefore an accurate theoretical treatment of the four-body scattering state
 is essential, which, however, is a hard task.
The neutrino-$^4$He inclusive cross sections have been evaluated in a 
shell-model approach 
in ref.\cite{suzuki2006}, while an {\it ab-initio} calculation has been done
by making use of the Lorentz Integral Transformation (LIT) method\cite{gazit2007}.
Also, a recent work of strength function (SF) based on the correlated Gaussians 
and the complex scaling method\cite{horiuchi2013}
has been applied to the neutrino reaction\cite{Murata}.

Among those methods, the LIT method has been widely applied to 
the break-up reactions of few-nucleon systems, 
where function $L(\sigma_R,\sigma_I)$ defined with the 
SF $R(\omega)$ ($\omega$: excitation energy) by an integral transformation,
\begin{eqnarray}
 L(\sigma_R,\sigma_I) = \int d\omega
  \frac{R(\omega)}{(\omega-\sigma_R)^2+\sigma_I^2},
\end{eqnarray}
plays a central role. The function $L$ can be calculated by 
using bound state like wave functions of the many body system.
Then the SF $R(\omega)$ is obtained by the inverse transformation 
of the above integral. Though the inversion of function expressed
by the convolution is in principle possible using Fourier 
transformation, the method used in the literature is  
$\chi^2$-fitting by using  assumed functional form of SF such as a sum
of exponentials \cite{Efros2007, Leidemann2008}.
However, the $\omega$-dependence of SF is not known a priori 
and the assumed functional form might lead to a false SF even the $\chi^2$ minimum
is achieved.  
In this report, we examine the maximum entropy method (MEM) as a new tool
for the inversion of the LIT.
In the MEM, we do not need to assume any functional form of the SF.
The MEM is widely used in the fields of the condensed-matter physics,
the Lattice QCD \cite{R.N.SilverD.S.Sivia1990, Gubernatis1991, Asakawa2001},
and the Green's functions Monte Carlo method \cite{Lovato2015}.

In section 2 and 3, we briefly explain the LIT method
and the MEM. In section 4, we apply the MEM to the inversion
of LIT. The results are reported for  spin-dipole SF
of $^4$He as an example.

\section{Lorentz Integral Transformation Method}
We briefly explain the LIT
method for calculating nuclear SFs.
The SF of transition operator $O$ is given as
\begin{eqnarray}
 R(\omega) = \Sigma_f \left|\bra{\psi_f}O
\ket{\psi_0}\right|^2 \delta(E_f-E_0-\omega),
\end{eqnarray}
where $\psi_0$ and $\psi_f$ are the initial and final states, respectively. Both the initial and final states are eigenstates of a Hamiltonian $H$ with the energies $E_0$ and $E_f$.
Here we define the LIT of SF as
\begin{eqnarray}
 L(\sigma_R,\sigma_I) = \int d\omega
  \frac{R(\omega)}{(\omega+E_0-\sigma_R)^2+\sigma_I^2}.
\label{lit-def}
\end{eqnarray}
From the definition of SF, the LIT can be written as
\begin{eqnarray}
 L(\sigma_R, \sigma_I)
&=& \bra{\psi_0}O^\dagger
  (H+E_0-\sigma_R+i\sigma_I)^{-1}(H+E_0-\sigma_R-i\sigma_I)^{-1}O\ket{\psi_0}\\
&=& \braket{\tilde{\psi}|\tilde{\psi}},
\end{eqnarray}
where
\begin{eqnarray}
 \ket{\tilde{\psi}} = (H+E_0-\sigma_R- i \sigma_I)^{-1}O\ket{\psi_0}.
\label{eq-psi}
\end{eqnarray}
We see that the norm of $\ket{\tilde{\psi}}$ is finite 
and thus $\ket{\tilde{\psi}}$ can be treated like a bound state, which is a great advantage
of this method, otherwise one has to construct the scattering state of a few-nucleon system.
In the LIT method, one at first calculates $L(\sigma_R,\sigma_I)$ and then
SF $R(\omega)$ is obtained by the inverse transformation of Eq. (\ref{lit-def}).

\section{Maximum Entropy Method}
We briefly explain how the MEM  \cite{MEM} can be used
to extract  the SF $R(\omega)$, knowing only
a set of $L(\sigma_R,\sigma_I)$  obtained from
the many-body calculation.
According to the Bayes' theorem \cite{bayes}, the most plausible 
SF ($R^{\rm MEM}(\omega)$) is given by the functional integral of the 
SF $R(\omega)$:
\begin{eqnarray}
 R^{\rm MEM} & = & \int [dR] R P[R|\bar{L},m],
\end{eqnarray}
where $P[R|\bar{L},m]$ is a conditional probability of SF $R$ 
for a given LIT denoted as $\bar{L}$ and 
prior information, denoted by $m$, for the SF.
$m$ is called the default model of the SF.
Introducing an auxiliary variable $\alpha$, we can rewrite the above formula as
\begin{eqnarray}
 R^{\rm MEM} & = & \int d\alpha  R^\alpha P[\alpha|\bar{L},m], \label{eq-smem}
\end{eqnarray}
where $R^\alpha$ depends on $\alpha$ in addition to our inputs $\bar{L}$ and $m$,
 and is given by
\begin{eqnarray}
 R^\alpha & = & \int [dR]  R P[R|\alpha,\bar{L},m] 
            \propto  \int [dR]  R P[\bar{L}|R,\alpha,m] P[R|\alpha,m].
\end{eqnarray}
Here $P[\bar{L}|R,\alpha,m]$ and  $P[R|\alpha,m]$ are called the likelihood 
function and prior probability, respectively and are given by
\begin{eqnarray}
P[\bar{L}|R,\alpha,m]  =  \frac{1}{Z_\chi}\exp\left(-\frac{1}{2}\chi^2\right),\ \   P[R|\alpha,m] = \frac{1}{Z_S} \exp\left(\alpha S\right).
\end{eqnarray}
with
\begin{eqnarray}
  \chi^2 & =& \Sigma_{l=1}^{N_{\sigma_R}}\frac{(\bar{L_l}-L_l)^2}{\delta_l^2},\ \   
  S = \Sigma_i^{N_\omega} \left(R_i - m_i - R_i \ln\frac{R_i}{m_i} \right)\\
  Z_\chi &=& \Pi_{l=1} \sqrt{2\pi\delta_l^2},\ \ 
  Z_S = \left(\frac{2\pi}{\alpha}\right)^{N_\omega/2}. 
\end{eqnarray}
Here $S$ is called the Shannon-Jaynes entropy. 
The excitation energy $\omega$ is discretized as $\omega_i$ ($i=1,2,...,N_\omega$) 
with equal spacing $\Delta_\omega$.
$R_i$ and $m_i$ are given by $R_i=R(\omega_i)\Delta_\omega$ 
and $m_i = m(\omega_i)\Delta_\omega$. 
$\bar{L}_l$ is the given $L$  at $\sigma_R = \sigma_R^l$
($l=1,2,...,N_{\sigma_R}$) with the error $\delta_l$.
$L_l$ is calculated from $R_i$ using Eq. (3) at $\sigma_R = \sigma_R^l$.
Combining the likelihood function and the prior probability,
the SF $R^\alpha$ is chosen  to maximize the probability of
$\displaystyle P[R|\alpha,\bar{L},m] \propto e^{Q(R)}$
with $Q(R) = \alpha S - \frac{1}{2}\chi^2$.

Finally, as in Eq. (8), $R^\alpha$ is convoluted with
the probability $P[\alpha|\bar{L},m]$ that has a sharp peak 
as a function of $\alpha$ and is written as
\begin{eqnarray}
P[\alpha|\bar{L},m] & = & \int [dR] P[R, \alpha|\bar{L},m]
  \propto P[\alpha|m] \int [dR] \frac{1}{Z_S Z_\chi}e^{Q(R)}.
\end{eqnarray}
Assuming that $P[\alpha|m]$ is constant,
% and replacing the functional intergration $[dR]$ by $\prod_i dR_i/\sqrt{R_i}$,
we obtain the SF $R^{\rm MEM}$ by integrating with respect to $\alpha$ around 
the sharp peak of $P[\alpha|\bar{L},m]$ as in Eq. (\ref{eq-smem}).

\section{Application of MEM to Inversion of LIT}

Our question is whether the MEM described 
in the previous section is useful to invert the integral transformation in Eq. (3).
For this purpose, we start from pseudo data of LIT, $L(\sigma_R, \sigma_I)$, that are generated from a 'known' SF $R^{\rm orig}(\omega)$
by using Eq. (3). We then use the MEM to obtain SF from the pseudo LIT data
without assuming any functional form for the 'reconstructed' $R(\omega)$.
We use the SF of $^4$He for the spin-dipole operator as $R^{\rm orig}$; this SF has been calculated in Ref. \cite{horiuchi2013}.
The SF for the spin-dipole operator that can induce neutrino and 
anti-neutrino reactions
($\pm$) is given by summing the final scattering states $\ket{\psi_f}$ as
\begin{eqnarray}
 R^{J\pm}(\omega) &=& \Sigma_f \left| \bra{\psi_f}\Sigma_j
 \left[{\mib \rho}_j\otimes{\mib \sigma}_j\right]_{(J)}\tau^\pm_j\ket{\psi_0}\right|^2
 \delta(E_f-E_0-\omega)
\end{eqnarray}
where $\Sigma_j$ denotes sum of the nucleons, ${\mib \rho}_j$, ${\mib \sigma}_j$ 
and $\tau_j$ are the internal coordinate, spin, and isospin of the {\it j}-th nucleon, respectively.
$\psi_0$ is the ground state of $^4$He.

\begin{figure}[H]
\includegraphics[width=10cm]{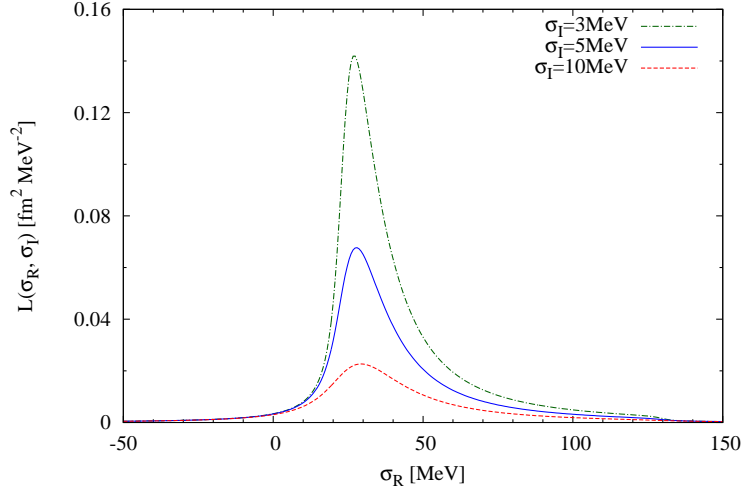}
\caption{Pseudo LIT data of $^4$He spin-dipole SF. 
The green dot-dashed, the blue solid and the red dotted curves 
show $L(\sigma_R, \sigma_I)$ for $\sigma_I$=3, 5, 10 MeV, respectively.}
\label{fig:lit-data}
\end{figure}

The pseudo LIT data $L(\sigma_R, \sigma_I)$ as a function 
of $\sigma_R$ are shown  in Fig. \ref{fig:lit-data}
for $\sigma_I=3, 5, 10$ MeV. The peak of $R^{\rm orig}$ becomes broader
in $L(\sigma_R, \sigma_I)$ as $\sigma_I$ is increased.

\begin{figure}[H]
\begin{minipage}{0.5\hsize}
\begin{center}
\includegraphics[width=8cm]{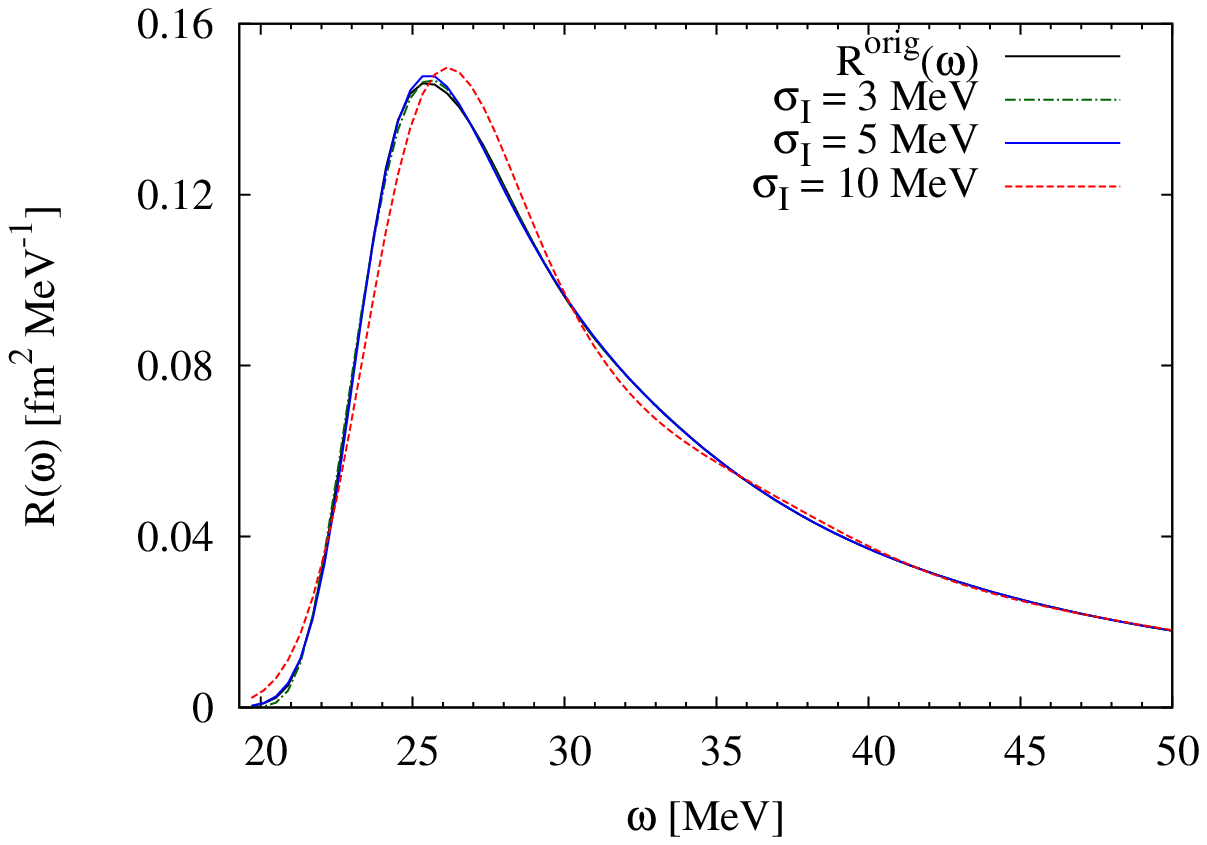}
\end{center}
\end{minipage}
\begin{minipage}{0.5\hsize}
\begin{center}
\includegraphics[width=8cm]{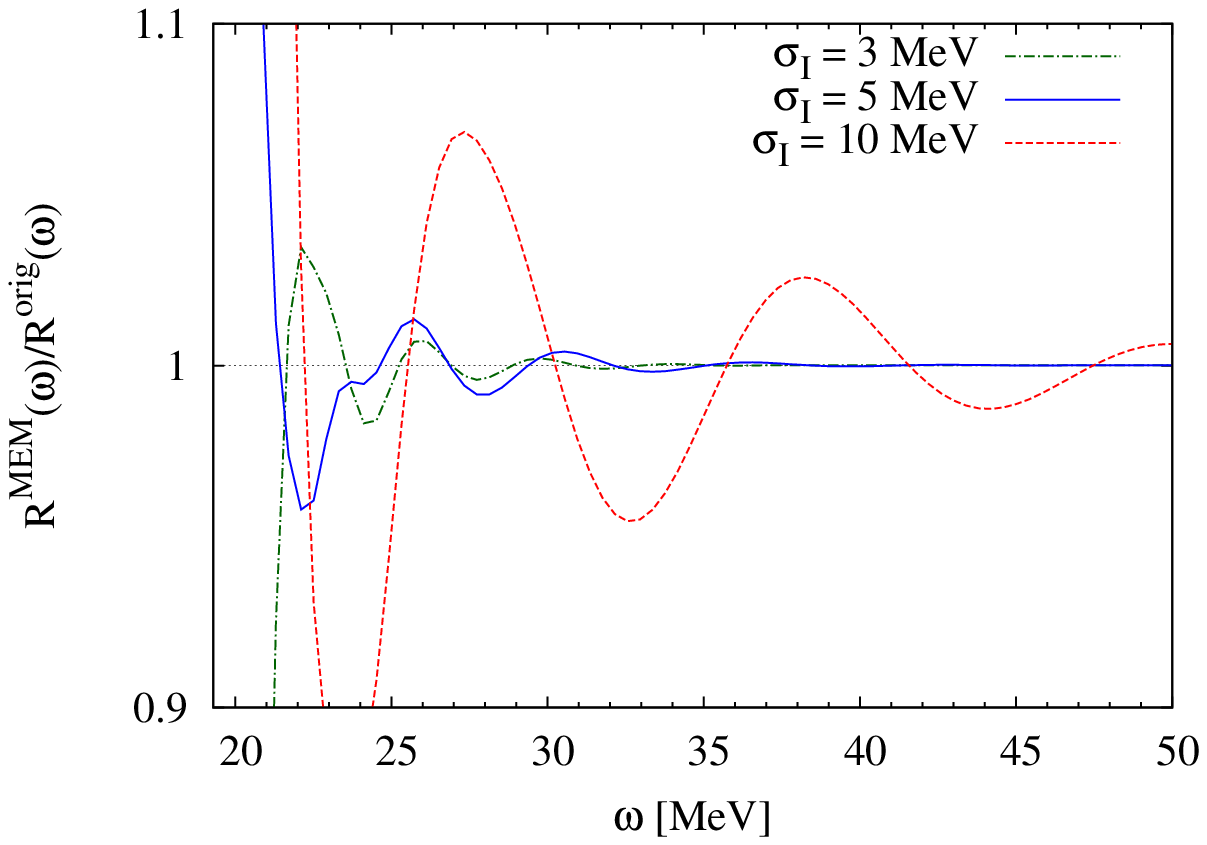}
\end{center}
\end{minipage}
\caption{The reconstructed SF $R^{\rm MEM}$(Left) and
the ratio  $R^{\rm MEM}/R^{\rm orig}$(Right).}
\label{fig:sf-mem}
\end{figure}

We then apply the MEM to those pseudo data $L(\sigma_R, \sigma_I)$ to obtain the $R^{\rm MEM}$.
Here we adopt a constant SF as the default model that is $m_i$ in Eq. (11).
The obtained $R^{\rm MEM}$ in comparison with the original SF $R^{\rm orig}$ are
shown in Fig. \ref{fig:sf-mem}. The left panel shows the $R^{\rm MEM}$ reconstructed 
from $L(\sigma_R, \sigma_I)$ with $\sigma_I$=3, 5, 10 MeV.
The right panel shows the ratio $R^{\rm MEM}/R^{\rm orig}$, showing deviation
of the reconstructed SF from the original one.
For $\sigma_I=3, 5$ MeV, 
the reconstructed SF $R^{\rm MEM}(\omega)$ agree well with $R^{\rm orig}(\omega)$
 except near the threshold.
The deviation of the ratio  $R^{\rm MEM}/R^{\rm orig}$ from one is 
within 1\% in the region $25$ MeV $\leq \omega \leq$ $120$MeV.
On the other hand, for $\sigma_I=$10 MeV, the peak of
$R^{\rm MEM}(\omega)$ shifts by 1 MeV, and $R^{\rm MEM}$ shows an oscillatory behavior for higher $\omega$ region.
The deviation from $R^{\rm orig}$ is more than 5\%, and is nonnegligible for $\omega <$ 30 MeV.

We have shown that the MEM can be successfully applied to the inversion of LIT
without assuming any functional form for SF $R(\omega)$,
provided we use LIT data with $\sigma_I$ sufficiently smaller than 10 MeV.
In general, the structures of $R(\omega)$ 
narrower than $\sigma_I$ are smeared out in $L(\sigma_R, \sigma_I)$,
and it is very hard to reconstruct the narrow structure of $R(\omega)$.
In our example, the width of the peak structure  of $R(\omega)$ is about 10 MeV.

In this work, we used the known SF, $R^{\rm orig}(\omega)$, to generate the pseudo data, $L(\sigma_R,\sigma_I)$. Thus, the pseudo data are very accurate at any $\sigma_R$ and $\sigma_I$.  In practice, however, we need obtain $L(\sigma_R,\sigma_I)$ from a discrete spectrum that is calculated with the eigenvalue method or the Lanczos algorithm \cite{Efros2007}. Therefore, more study is needed to examine if the MEM method applied to the LIT inversion works in the practical situations.

\end{document}